\documentclass[aps,prl,twocolumn,showpacs]{revtex4}
\usepackage{amssymb}
\usepackage{amsmath, amsthm}
\newcommand{\be}{\begin{equation}}
\newcommand{\ee}{\end{equation}}

\begin{document}


\title{The Jebsen-Birkhoff theorem in alternative gravity} 


\author{Valerio Faraoni}
\email[]{vfaraoni@ubishops.ca}
\affiliation{Physics Department, Bishop's University\\
Sherbrooke, Qu\'ebec, Canada J1M~1Z7
}

\begin{abstract} 
We discuss the validity, or lack thereof, of the Jebsen-Birkhoff 
theorem in scalar-tensor theories 
by generalizing it  and regarding the 
Brans-Dicke-like scalar as effective matter. Both the 
Jordan and Einstein frames are discussed and an apparent  
contradiction between  static spherical solutions of 
scalar-tensor gravity and Hawking's theorem on Brans-Dicke black 
holes is clarified. The results are applied to metric and 
Palatini $f(R)$ gravity.
\end{abstract}


\maketitle

\section{Introduction}

The recent attempts to explain the acceleration of the universe 
discovered with type Ia supernovae \cite{SN} have led to the 
introduction of an {\em ad hoc} dark energy comprising 76\% of 
the energy content of the universe and with an extreme (possibly 
phantom) equation of state $P\simeq -\rho$, where $\rho$ and $P$ 
are the effective energy density and pressure of the dark 
energy \cite{Linderresletter}. Displeased with this explanation, 
many  authors have turned to the possibility that dark energy 
does not  exist and, instead, gravity must be modified in the 
infrared  sector \cite{CCT, CDTT} and the so-called $f(R)$ 
gravity theories have been resurrected for this purpose. $f(R)$ 
gravity theories in use in cosmology  come in three 
versions, called the  metric formalism \cite{CCT, CDTT}, Palatini 
formalism \cite{Vollick}, and 
metric-affine gravity \cite{metricaffine} (see  \cite{review} for  
a recent review and \cite{otherreviews} for shorter  
introductions).  The study of spherically symmetric solutions 
has been crucial to understand the weak-field limit of these 
theories and 
confront them with Solar System observations (remember that the 
three classical tests of General Relativity are based on the 
spherically symmetric Schwarzschild solution \cite{Wald, Will}). 
Morever, the study  of strong field instabilities potentially 
fatal for the theory  is carried out in spherical symmetry 
\cite{Frolovetal}. It is 
therefore important  to understand how the  Jebsen-Birkhoff 
theorem well known from 
spherical symmetry in General Relativity generalizes, or fails, 
in these extensions of Einstein's theory. 
Students beginning in General Relativity are familiar 
with Birkhoff's theorem stating that a spherically symmetric 
solution of the vacuum Einstein equations is static  
\cite{Birkhoff} (it appears  \cite{DeserFranklin, 
JohansenRavndal, Deser2005} that this result 
had  actually been discovered  by Jebsen two years before 
Birkhoff    \cite{Jebsen}).  It is now well-known 
that the Jebsen-Birkhoff  theorem does not hold in metric $f(R)$ 
gravity  while it is valid in Palatini 
$f(R)$  gravity \cite{review}.  Since both metric and Palatini 
$f(R)$ 
gravity  admit a description  as special scalar-tensor theories 
(Brans-Dicke theories with values of the Brans-Dicke 
parameter $\omega=0$ and $\omega=-3/2$, respectively, and endowed 
with a special 
potential for the Brans-Dicke scalar \cite{STequivalence, 
review}), a complete understanding of the Jebsen-Birkhoff theorem 
in $f(R)$ gravity 
requires its understanding  in general 
scalar-tensor gravity, which is the goal of the present paper. 
Beginning with the realization that the Brans-Dicke-like scalar 
field of scalar-tensor gravity can, and most often is, regarded
as an effective form of matter for effective Einstein equations, 
one is led to formulate a generalized  Jebsen-Birkhoff 
theorem in the  presence of matter in General Relativity, which 
is convenient for the  study of spherical symmetry in 
scalar-tensor 
gravity  (this is done in Secs.~II and~III using the Jordan 
frame).  An alternative approach to scalar-tensor gravity 
consists of using its Einstein frame formulation, 
which is employed in Sec.~IV. In this regard, an apparent 
contradiction between certain long-known exact spherical 
solutions of scalar-tensor theory \cite{Brans, CampanelliLousto}  
and Hawking's theorem for Brans-Dicke black holes 
\cite{HawkingBD} is elucidated. Almost as a byproduct of this  
work, light is shed on the  validity, or lack thereof, of the 
Jebsen-Birkhoff theorem of  Palatini and metric $f(R)$ gravity 
in Sec.~V.

We adopt the notations of Ref.~\cite{Wald}. The cosmological 
constant $\Lambda$, if  present, is seen as an effective form of 
matter described by the  formal stress-energy tensor 
$T_{ab}^{(\Lambda)}=-\, \frac{\Lambda}{8\pi G} \, g_{ab}$ (that 
is, in the following, ``vacuum'' implies that $\Lambda=0$).

\section{The Jebsen-Birkhoff theorem in General Relativity}

The most general spherically symmetric line element can be 
written as
\be \label{metric}
ds^2=-A^2(t, r) dt^2+B^2(t, r) dr^2 +r^2 d\Omega_2^2 \,,
\ee
where $d\Omega_2^2= d\theta^2 +\sin^2 
\theta \, d\varphi^2  $ is the line element on the unit 
2-sphere  and $r$ is an areal radius. The metric components 
 $g_{0i} $ 
($i=1,2,3$), if present, can be eliminated by redefining the 
coordinates $t$ and $r$ (\cite{Jebsen}, see 
also \cite{LandauLifsits}).  The $ \left(0,1\right), 
 \left(0,0\right), \left(1,1\right)$,  $\left(2,2\right)$, and  
$\left(3,3\right)$ components of the  Einstein field 
equations 
\be\label{efe}
R_{ab}-\frac{1}{2}\, g_{ab}R=8\pi G \, T_{ab}^{(m)} \,,
\ee
where $T_{ab}^{(m)}$ is the matter energy-momentum tensor, yield
\begin{eqnarray}
&&  \frac{\dot{B}}{B}=4\pi G r \, T_{01}^{(m)} \,, \label{efe01} 
\\
&& \nonumber \\
&&   \frac{1}{r^2} + \frac{2B'}{B^3 r}-\frac{1}{B^2 
r^2}  = 8\pi G \, \frac{T_{00}^{(m)} }{A^2} \,, \label{efe00} \\
&& \nonumber \\
&&   \frac{2A'}{Ar} -\frac{B^2}{ r^2}+ \frac{1}{r^2} 
=8\pi G \, T_{11}^{(m)} \,, \label{efe11} \\
&& \nonumber \\
&&   \frac{r}{ B^3 } \left(  \frac{A'B}{A} -B' -\frac{r B^2 
\ddot{B} }{A^2} + \frac{r\dot{A} B^2 \dot{B}}{A^3} - 
\frac{rA'B'}{A} +
\frac{r A'' B}{A} \right) \nonumber\\
&&\nonumber\\
&& =8\pi G \, T_{22}^{(m)}   \,, \label{efe22}\\
&&\nonumber\\
&&   \frac{r\sin^2\theta }{ B^3 } \left(  \frac{A'B}{A} -B' 
-\frac{r B^2 
\ddot{B} }{A^2} + \frac{r\dot{A} B^2 \dot{B}}{A^3} - 
\frac{rA'B'}{A} \right. \nonumber\\
&&\nonumber\\
&&  \left. + \frac{r A'' B}{A} \right) =8\pi G \, T_{33}^{(m)}  
\,, \label{efe33}
\end{eqnarray}
where a prime and an overdot denote differentiation 
with respect to $r$ and $t$, respectively.

\subsection{The non-vacuum case}

Let us consider a timelike observer at rest in the coordinate 
system $\left( t,r,\theta, \varphi \right)$ adapted to the 
spherical symmetry of the metric~(\ref{metric}), {\em i.e.}, one 
with 4-velocity  $ u^{\mu}=\left( A^{-1}, 0,0,0 \right)$. The 
matter energy density relative to this observer ``at rest'' is
\be
\rho \equiv T_{ab}^{(m)} u^a u^b =\frac{T_{00}^{(m)} }{A^2} 
\ee
and coincides with the right hand side of the 
Hamiltonian constraint~(\ref{efe00}) 
apart from the factor $8\pi G$. The radial energy 
current relative to this observer is
\be
J_{(r)} \equiv - T_{ab}^{(m)} \, u^a 
\, e^b_{(r)}=\frac{T_{01}^{(m)} }{AB} \,, 
\ee
where $e^b_{(r)}$ is the spacelike unit vector in the radial 
direction with components $e^{\mu}_{(r)}=\left( 0, B^{-1}, 0, 0 
\right)$. Finally, the radial pressure relative to this observer 
is 
\be
P_{(r)} \equiv T_{ab}^{(m)} e^a_{(r)} \, e^b_{(r)} = 
\frac{T_{11}^{(m)} }{B^2} \,.
\ee
The non-radial stresses $T_{ij}^{(m)} $ with $i\neq j $ 
($i,j=1,2,3$)  vanish identically because of the Einstein 
equations and the fact 
that the components $G_{ij}$ of the Einstein tensor with $i \neq 
j$ vanish in spherical symmetry.  Furthermore, 
eqs.~(\ref{efe22}) and 
(\ref{efe33}) imply that $ 
\frac{T_{33}^{(m)}}{\sin^2\theta}=T_{22}^{(m)}$. 

The Einstein equations require that the matter distribution be  
spherically symmetric, {\em i.e.}, that the derivatives with 
respect to $\theta$ and $\varphi$ of $ \rho, J_{(r)}$, and 
$P_{(r)}$ vanish. A spherically symmetric $T_{ab}^{(m)} $ is 
said to describe a {\em static matter distribution} 
iff \footnote{Alternatively, one can require that the Lie 
derivatives of $T_{ab}^{(m)}$ along the directions of the two 
Killing vectors $t^a$ (timelike) and $\psi^c$ (spacelike) vanish 
\cite{SchleichWitt}.}
 \be
A u^c\nabla_c \rho = \frac{\partial \rho}{\partial t}= 0\,, 
\;\;\;\;
A u^c\nabla_c P_{(r)}=\frac{\partial P_{(r)} }{\partial t}=0 \,, 
\;\;\;\;  J_{(r)}=0   \,.
\ee
Eq.~(\ref{efe01}) with the assumption $J_{(r)}=0$ (equivalent to 
$T_{01}^{(m)} =0$ in a region where $A$ and $B$ are finite and 
positive) 
guarantees that $\dot{B}=0$, {\em i.e.}, $B=B(r)$. 
Then eq.~(\ref{efe11}) with the assumption $\partial 
P_{(r)}/\partial t=0$ guarantees that $A'/A$ is 
time-independent, which only leaves the possibility that $A$ 
depends on time through a multiplicative factor, $A(t,r)=f(t) 
a(r)$. But then the line element assumes the form
\be
ds^2=-a^2(r) f^2(t) dt^2+B^2(r) dr^2 +r^2 d\Omega_2^2 
\ee
and the redefinition of the time coordinate 
$t\rightarrow \bar{t}$ with $d\bar{t} \equiv f(t) 
dt$ 
then absorbs the factor $f(t)$ into $\bar{t}$ and 
casts the metric into locally static 
form. Then the Einstein equations~(\ref{efe22}) and 
(\ref{efe33})  imply that also $T_{22}^{(m)}$ and $ T_{33}^{(m)}$ 
are  
time-independent, and this is true for the tangential 
pressures
\begin{eqnarray}
P_{(\theta)} & \equiv & T_{ab}^{(m)} 
e^a_{(\theta)}e^b_{(\theta)}=\frac{T_{22}^{(m)} }{r^2}\nonumber\\
&&\nonumber\\
&=& P_{(\varphi)} \equiv T_{ab}^{(m)}  
\, e^a_{(\varphi)} \, e^b_{(\varphi)} 
=  \frac{T_{33}^{(m)} }{r^2\sin^2\theta} 
\end{eqnarray}
as well, where $e^a_{(\theta)}$ and $ e^a_{(\varphi)}$ are 
spacelike unit 
vectors in the angular directions. We have 
therefore the\\\\
{\bf Jebsen-Birkhoff theorem (version~1):} {\em If a solution of 
the Einstein equations is 
spherically symmetric and the matter distribution is static 
($ \frac{\partial \rho}{\partial t}= 
\frac{\partial P_{(r)} }{\partial t}=0 $ and $  J_{(r)}=0$),  
then the metric is static in a region in which $t$ remains  
timelike and $\left( r, \theta, \varphi \right) $ stay
spacelike.}\\

The restriction to the region where the coordinates preserve 
their timelike or spacelike character is necessary for the 
validity of the theorem, as originally noted by Ehlers and 
Krasinski 
\cite{EhlersKrasinski}. For example, this restriction is not 
satisfied in the region inside the Schwarzschild black 
hole horizon or outside the de Sitter cosmological horizon, where 
these metrics become time-dependent.  A better statement of the 
Jebsen-Birkhoff theorem is that, under the conditions stated ``a 
spherically symmetric solution admits, besides the SO(3) 
generators, an additional hypersurface-orthogonal Killing vector 
field'' (\cite{EhlersKrasinski}, see also \cite{Exact, 
FrolovNovikov, SchleichWitt}). However, the metric is static 
only where this additional Killing field remains timelike, which 
excludes the black hole horizon (if this exists  and coincides  
with the Killing horizon) where the Killing field  goes 
through zero to change sign in the interior \footnote{The local 
nature of the Jebsen-Birkhoff theorem is emphasized in 
\cite{SchleichWitt}.}.  
The theorem 
does not make statements valid  on a black hole horizon surface 
which may 
be  present and  it does not imply that the solution of the 
Einstein  equations is the Schwarzschild metric. A 
solution  with a matter distribution diverging on a horizon is 
possible; this happens, for example, for the Brans class~I 
solutions of 
scalar-tensor 
gravity \cite{Brans}  (in which the  Brans-Dicke-like 
scalar  field can be  
regarded as a form of effective matter) 
which are static and  spherically symmetric but have a scalar 
field diverging at the  horizon---see Sec.~IV.

It might seem that the assumptions on the matter $T_{ab}^{(m)}$ 
are too 
strong because this tensor contains the metric $g_{ab}$ and, 
therefore, any assumption on $T_{ab}^{(m)}$ means to require 
already,  in some way, that the metric $g_{ab}$ is static. 
Indeed, this 
is 
true but the assumption of a static matter distribution 
leaves room for physically relevant  situations. The first is 
vacuum, in which no-matter is necessarily (and trivially) of the 
static 
form  
and gives rise to the more familiar version of the 
Jebsen-Birkhoff 
theorem upon which we comment in the next subsection. 
Another 
non-trivial situation is that of a cosmological constant seen as 
a form of effective matter described by $T_{ab}^{(\Lambda)}$, 
which is spherically symmetric and static. In this case  the 
solution  is not  the Schwarzschild but the 
Schwarzschild-(anti)de 
Sitter one. It is well known that, for example, the 
Schwarschild-de Sitter (Kottler) line element  can be put in the 
locally static form 
\begin{eqnarray}
ds^2 &= & -
\left( 1-\frac{2GM}{r}-\frac{\Lambda r^2}{3} \right) dt^2 
\nonumber\\
&&\nonumber\\
& + &  \left( 1-\frac{2GM}{r}-\frac{\Lambda r^2}{3} \right)^{-1} 
dr^2  +r^2 d\Omega_2^2  
\end{eqnarray}
in the region of spacetime manifold between the black hole 
and the cosmological horizons. Another physical situation 
allowed by the assumptions of the 
theorem is electro-vacuum \cite{Das60}. If the mass distribution 
carries  a  static electric charge $Q$ with no radial current the 
solution is the static Reissner-Nordstrom metric given by
\begin{eqnarray}
ds^2 & = & -
\left( 1-\frac{2GM}{r}+ \frac{Q^2 }{r^2} \right) dt^2 \nonumber\\
&&\nonumber\\
& + &  \left( 1-\frac{2GM}{r} + \frac{Q^2}{r^2} \right)^{-1} 
dr^2  +r^2 d\Omega_2^2  \,.
\end{eqnarray}

The absence of a radial energy current, 
$J_{(r)}=0$, without the 
assumption of staticity of $\rho$ and $P_{(r)}$  is not 
sufficient  to 
guarantee that the metric is static. A counterexample is the 
McVittie solution describing a spherical object 
embedded in a cosmological background \cite{McVittie} for which 
$J_{(r)}=0$  but the metric 
is time-dependent except for the special case in 
which it reduces to the Schwarzschild-(anti)de Sitter one.

Although less familiar than the vacuum version, version~1 of 
the Jebsen-Birkhoff theorem is more suitable for discussing 
the spherically symmetric solutions of scalar-tensor gravity 
because the Brans-Dicke-like scalar field present in these 
theories acts as an effective form of matter.

\subsection{The vacuum case}

By regarding vacuum ($T_{ab}^{(m)} =0$)  as a form of static 
matter, the Jebsen-Birkhoff theorem assumes its most 
familiar 
form.\\\\
{\bf Jebsen-Birkhoff theorem (version~2):} {\em a  
spherically symmetric solution of the vacuum Einstein equations  
is necessarily static in a region in which $t$ remains  
timelike and $\left( r, \theta, \varphi \right) $ stay
spacelike.}\\

The vacuum assumption rules out the possibility of a  
cosmological constant as well as 
electro-vacuum, and the Schwarzschild-(anti)de Sitter and 
Reissner-Nordstrom solutions. The solution is then forced to be 
Schwarzschild.

Note that the vacuum as defined by $T_{ab}=0$ is not necessarily 
a trivial configuration in alternative theories of gravity. In 
scalar-tensor gravity  the 
Brans-Dicke-like scalar field $\phi$ describing the 
gravitational field together with the metric $g_{ab}$ may be  
non-constant and still give a vanishing effective stress-energy 
tensor $T_{ab}^{( \phi)}$; this is referred to 
as a ``non-gravitating'' or ``stealth'' 
scalar field. Two  examples of massive waves of  
a $\phi$-field coupled nonminimally to the Ricci curvature and 
with a potential are given in \cite{Beatoetal05}; they achieve 
$T_{ab}^{(\phi)}=0$ and no other form of matter is present. As a 
result, the spacetime is Minkowskian (spherically symmetric and 
static), providing a non-trivial realization of ``vacuum'' and of 
version~2 of the Jebsen-Birkhoff theorem. Other examples of 
non-gravitating matter distributions are given in 
Refs.~\cite{otherstealth}.

In the language of field theory, the field content of General 
Relativity consists only of a  massless spin two field and 
gravitational 
radiation is 
quadrupole to lowest order. Hence spherically symmetric sources 
cannot excite gravitational radiation and the 
gravitational charge ({\em i.e.}, the mass-energy) of the source
is conserved---the spacetime around a spherically symmetric 
source 
must be static.

It is usually remarked that the Jebsen-Birkhoff theorem allows, 
as a corollary, an extension to General Relativity of the 
iron sphere theorem  of Newtonian gravity stating that the 
gravitational 
field of a spherically symmetric distribution of mass inside a 
spherical cavity is zero. In spherical symmetry, if the energy 
distribution inside a cavity is static, the solution of the 
Einstein equations will be static. In vacuo, it is Minkowski 
space, {\em i.e.}, the Schwarzschild solution corresponding to 
zero mass.  If matter inside the cavity consists  only of a 
cosmological  constant the interior solution is spatially 
homogeneous and isotropic about every point and, 
therefore, (anti-)de Sitter.

\section{The Jebsen-Birkhoff theorem in scalar-tensor gravity}

With the advent of the Jordan \cite{Jordan} and Brans-Dicke 
\cite{BD} theories first and of 
scalar-tensor theories later \cite{ST}, 
the validity of the Jebsen-Birkhoff 
theorem was investigated in alternative gravity   
\cite{Schucking57, OHanlonTupper72, Reddy73,  Reddy77, 
KroriNandy77, SinghRai79, DuttaBatta80, Singh86, Singh86again, 
Reddyetal87, 
Reddy88, BuddyReddy89}.

In  general, the theorem does not hold in scalar-tensor gravity: 
one needs to impose that the effective stress-energy tensor 
$T_{ab}^{(\phi)}$ of the Brans-Dicke-like scalar field of 
the  theory is time-independent in order for the metric to be 
static (this requirement is usually achieved by imposing that 
$\phi$ is time-independent, but is also obtained by a 
stealth field $\phi$ \cite{Beatoetal05, otherstealth}). The 
failure of the 
theorem in the presence 
of time-dependent scalars opens the door for new phenomenology in 
scalar-tensor gravity which is  unknown in General Relativity. 
The failure of the Jebsen-Birkhoff theorem is to be expected: 
since scalar-tensor gravity has a new spin zero degree of freedom 
in comparison with General Relativity, scalar monopole radiation 
can occur. In Einstein's theory monopole radiation is 
forbidden by the Jebsen-Birkhoff theorem, which is a consequence 
of the  fact that the gravitational field is represented only by 
a spin two field. Since in 
Einstein's theory gravitational 
radiation is necessarily quadrupole to lowest order, spherically 
symmetric pulsating sources cannot generate gravitational 
radiation and the metric must be static. This is no longer true 
in 
scalar-tensor gravity, in which the time-varying monopole 
moment of a radially pulsating spherical source  generates 
propagating scalar radiation which makes also the metric 
non-static.

Let us consider a scalar-tensor theory of gravity described by 
the action
\begin{eqnarray} 
S_{ST} & = & \frac{1}{16\pi}\int d^4x \, \sqrt{-g} \left[ \phi R 
-\frac{\omega(\phi)}{\phi} \, g^{ab}\nabla_a \phi  \,
\nabla_b \phi -V(\phi) \right] \nonumber\\
 &+ & S^{(m)} \,.  \label{STaction}
\end{eqnarray}
The field equations can be written in the form of effective 
Einstein equations as
\begin{eqnarray}
 R_{ab}-\frac{1}{2}\, g_{ab}R & = & \frac{8\pi}{\phi} \, 
T_{ab}^{(m)} \nonumber\\
&&\nonumber\\
& + & \frac{\omega(\phi)}{\phi^2} \left( 
\nabla_a \phi 
\nabla_b \phi -\frac{1}{2}\, g_{ab} \nabla^c \phi
\nabla_c \phi \right) \nonumber\\
&&\nonumber\\
& + & \frac{1}{\phi} \left( \nabla_a 
\nabla_b   \phi -g_{ab}\Box \phi \right) - \frac{V(\phi)}{2\phi} 
\, g_{ab}  \nonumber\\
&&\nonumber \\
&\equiv & \frac{8\pi}{\phi} 
\left(T_{ab}^{(m)}+ T_{ab}^{(\phi)} \right) 
\,,\label{STfield1} \\
&&\nonumber \\
 \left( 2\omega+3 \right) \Box\phi &= & 8\pi 
T^{(m)}-\frac{d\omega}{d\phi}\, \nabla^c \phi \nabla_c \phi
+\phi\, \frac{dV}{d\phi} -2V \,, \nonumber\\ 
&& \label{STfield2}
\end{eqnarray}
where $T_{ab}^{(m)}$ is the matter energy-momentum tensor and we 
assume that $\phi>0$ in conjunction with $\omega>-3/2$ to 
guarantee the positivity of the effective  
gravitational coupling \cite{Nordtvedt}
\be
G_{eff}=\frac{2\left( \omega+2 \right)}{2\omega +3}\, 
\frac{1}{\phi} \,.
\ee
In this form, it is easy to see that the scalar field $\phi$ acts 
as an effective form of matter in the field 
equations~(\ref{STfield1}) and therefore, by imposing that 
the matter stress-energy tensor $T_{ab}^{(m)}$ vanishes, one is 
left with an  
effective stress-energy tensor $T_{ab}^{(\phi)} $ such that 
$T_{00}^{(\phi)}$ could be time-dependent and $T_{0i}^{(\phi)} 
\neq 0 $ if $\phi$ depends on time. In other words, a 
time-dependent Brans-Dicke-like field $\phi$ spoils the 
validity of the Jebsen-Birkhoff theorem and only the assumption 
that $\phi$ is time-independent (or that 
$T_{ab}^{(\phi)}$ is static, as for a time-dependent   
stealth field $\phi$) restores the staticity of a 
spherically symmetric solution. This can be checked explicitly 
using the field equations, which we do in the following.

\subsection{The trivial case $\phi=$constant}

The case  $\phi=$const.$ \equiv \phi_0 > 0 $ is trivial and  
eq.~(\ref{STfield1}) reduces to 
\be
R_{ab}-\frac{1}{2}\, g_{ab}R =\frac{8\pi }{\phi_0}\, 
T^{(m)}_{ab}-\frac{V_0}{2\phi_0}\, 
g_{ab} 
\ee
where $V_0 \equiv V(\phi_0)$, so that the theory degenerates to 
General Relativity with 
the   
cosmological  constant $\Lambda \equiv V_0/(2\phi_0)$. If 
$T_{ab}^{(m)}$ is such  that the  energy distribution is static 
(including the case  $T_{ab}^{(m)}=0$), version~1 of the 
Jebsen-Birkhoff theorem holds and the metric is static in the 
region in which the coordinate gradients preserve their causal 
character. The same happens if $T_{ab}^{( \phi)}$ vanishes with 
$\phi \neq$const.

\subsection{Static (but non-constant) Brans-Dicke-like field}

Let us assume that the spacetime metric which solves the field 
equations~(\ref{STfield1}) and (\ref{STfield2}) is spherically 
symmetric 
with line element~(\ref{metric}). Then 
\be\label{gradgrad}
\nabla^c\phi \nabla_c\phi = -\frac{\dot{\phi}^2}{A^2}+ 
\frac{\phi'^2}{B^2} \,, 
\ee
the only  non-vanishing Christoffel symbols  are 
\begin{eqnarray}
&&  \Gamma^0_{00}=\frac{\dot{A}}{A} \,, \;\;\;\;\;\;
\Gamma^0_{01}=\Gamma^0_{10} = \frac{A'}{A} \,, \;\;\;\;\;\;
\Gamma^0_{11}=\frac{B\dot{B}}{A^2}   \\
&&\nonumber\\
&&  \Gamma^1_{00}=\frac{AA'}{B^2} \,, \;\;\;\;\;\;
\Gamma^1_{01}=\Gamma^1_{10} = \frac{\dot{B}}{B} \,, \;\;\;\;\;\;
\Gamma^1_{11}=\frac{B'}{B} \,,\\
&&\nonumber\\
&& \Gamma^1_{22}=- \frac{r}{B^2} \,, \;\;\;\;\;\;
\Gamma^1_{33}=- \frac{r}{B^2}\, \sin^2 \theta \,, \\
&& \nonumber\\
&& \Gamma^2_{12}=\Gamma^2_{21}=\frac{1}{r} \,, \;\;\;\;\;\;
\Gamma^2_{33}=-\sin\theta \cos\theta  \,, \\
&& \nonumber\\
&& \Gamma^3_{13}=\Gamma^3_{31}= \frac{1}{r}   \,, \;\;\;\;\;\;
\Gamma^3_{23}=\Gamma^3_{32}=\frac{\cos\theta}{\sin\theta} \, ,
\label{Christoffel}
\end{eqnarray}
and the d'Alembertian of $\phi$ is 
\begin{eqnarray}
\Box \phi & = &  -\, \frac{1}{A^2} \left( 
\ddot{\phi}-\frac{\dot{A}}{A} \, \dot{\phi} -\frac{AA'}{B^2} \, 
\phi' \right) \nonumber\\
&&\nonumber\\
& + &  \frac{1}{B^2} \left( \phi''-\frac{B\dot{B}}{A^2} \, 
\dot{\phi} 
-\frac{B'}{B}\, \phi' \right)+\frac{2\phi'}{rB^2} \,.\label{Box}
\end{eqnarray}
The  $ \left(0,1\right),   \left(0,0\right)$, 
and $ \left(1,1\right)$ 
components of the  field equations~(\ref{STfield1}) yield 
\begin{eqnarray}
&&  \frac{2\dot{B}}{Br}=\frac{8\pi}{\phi}\,   T_{01}^{(m)} 
+\omega \, \frac{ \dot{\phi}\phi' }{\phi^2}  +\frac{1}{\phi} 
\left( \dot{\phi}' -\frac{A'}{A} \, \dot{\phi} -\frac{\dot{B}}{B} 
\, \phi' \right) \,, \nonumber\\
&& \label{STfe01} \\
&&  A^2 \left( \frac{1}{r^2} +\frac{2B'}{B^3 r}-\frac{1}{B^2 r^2} 
\right) = \frac{ 8\pi }{\phi}\,  T_{00}^{(m)} \nonumber\\
&&\nonumber\\
&& + \frac{\omega }{2\phi^2} \left( \dot{\phi}^2 
 +\frac{A^2}{B^2} \, \phi'^2 \right) \nonumber \\
&&\nonumber \\
&& + \frac{A^2}{B^2 \phi} \left(  \phi''  
-\frac{B\dot{B}}{A^2} 
\, \dot{\phi} -\frac{B'}{B} \, \phi' +\frac{2\phi'}{r}  \right) 
+\frac{VA^2}{2\phi} \,, \label{STfe00} \\
&& \nonumber \\
&&   \frac{2A'}{Ar} -\frac{B^2}{ r^2}+ \frac{1}{r^2} 
=\frac{8\pi}{\phi} \,  T_{11}^{(m)}
+\frac{\omega}{\phi^2} 
\left(  \phi'^2 +\frac{B^2}{A^2} \, \dot{\phi}^2 \right)  \nonumber \\
&&\nonumber\\
&& +\frac{B^2}{A^2\phi} \left( 
\ddot{\phi}-\frac{\dot{A}}{A} \, 
\dot{\phi} -\frac{AA'}{B^2} \, \phi' -\frac{2A^2}{B^2 r} \phi' 
\right) -\frac{VB^2}{2\phi} \,. \nonumber\\
&& \label{STfe11} 
\end{eqnarray}
If there are no radial energy currents ($T_{01}^{(m)}=0$)  and, 
in particular, in vacuo ($T_{ab}^{(m)}=0$) 
and assuming that $\phi$ is 
time-independent ($\phi=\phi(r)$), eq.~(\ref{STfe01}) yields
\be
\frac{\dot{B}}{B}\left( \frac{2}{r}+\frac{\phi'}{\phi} \right)=0 
\,,
\ee
so that either $\dot{B}=0$ or $\phi(r)= C / r^2 $, where 
$ C> 0 $  is a constant. If $\phi = C / r^2 $ then eq.~(\ref{STfe00}) 
yields 
\be
B^2=\frac{2C \left( 2\omega+3 \right)}{2C+Vr^4} 
\ee
in vacuo and $B=B(r)$, therefore  we   
focus on  the $\dot{B}=0$ situation. Eq.~(\ref{STfe11}) reduces 
to
\begin{eqnarray}
\frac{A'}{A} \left( \frac{2}{r}+\frac{\phi'}{\phi} \right) 
& = &  \frac{8\pi}{\phi} \, T_{11}^{(m)} + \frac{ B^2-1 }{r^2}  
+\omega \left( 
\frac{\phi'}{\phi} \right)^2 -\frac{2}{r} \, \frac{\phi'}{\phi} 
\nonumber\\
&&\nonumber\\
&- & \frac{B^2 V}{2\phi} \,.
\end{eqnarray}
If $\partial T_{11}^{(m)} /\partial t = 0 $ (in particular in 
vacuo), the right  hand side  of this equation is 
$t$-independent, implying that $A$ can depend from $t$ at most 
through a 
multiplicative factor as in $A(t,r)=f(t)a(r)$. In this case a 
redefinition of the time coordinate $ d\bar{t}=f(t)dt$ absorbs 
the time-dependent factor into $\bar{t}$ and the metric can be 
cast in locally static form. Then, the other field equations 
imply 
that also the radial pressures $T_{ii}^{(m)}$ ($i=1,2,3$) are 
time-independent.

The fact that the scalar field $\phi$ needs to be static or  
non-gravitating in order 
to rescue the Jebsen-Birkhoff theorem was established on several 
occasions for particular scalar-tensor theories. 
Sch\"{u}cking \cite{Schucking57} derived the result for Jordan's 
theory, a precursor of Brans-Dicke theory \cite{Jordan};  Reddy 
\cite{Reddy77} studied the 
electro-vacuum case of the Sen-Dunn theory \footnote{The Sen-Dunn 
theory \cite{SenDunn} is a scalar-tensor theory in which the 
combination of second 
derivatives  $\nabla_a \nabla_b  \phi - g_{ab}\Box \phi$  
on the right hand side of the  field equations~(\ref{STfield1}) 
vanishes identically, making a massless $\phi$ satisfy the weak 
energy condition in the Jordan frame.}  
 and of  the  conformally  coupled scalar field  
\cite{Reddy88} (the former was revisited in \cite{DuttaBatta80}  
and the latter in \cite{Singh86}; the Sen-Dunn theory 
was reconsidered, and some errors of \cite{Reddy77} 
corrected, in \cite{KroriNandy77}).
Venkateswarlu and  Reddy studied electrovacuum in more general 
scalar-tensor theories \cite{BuddyReddy89}.

To summarize, when the Brans-Dicke-like scalar is static but 
non-constant, the solution of the field equations can be 
different from Schwarzschild-(anti)de Sitter. If this field is 
constant the  theory reduces to General Relativity, for which 
the Jebsen-Birkhoff theorem holds.  As a corollary, the 
spherically symmetric solution 
inside an  empty cavity is  static only if the scalar field 
$\phi$ is assumed to be static  or non-gravitating (Minkowski 
space if there is no potential and 
(anti-)de Sitter  space if the potential is non-zero).

\section{The Jebsen-Birkhoff theorem in the Einstein 
frame representation of scalar-tensor gravity }

By performing the conformal transformation of the metric and 
redefining non-linearly the Brans-Dicke-like scalar as in 
\begin{eqnarray}
&& g_{ab}\rightarrow \tilde{g}_{ab}=\Omega^2 \, g_{ab}, 
\,\,\,\,\,\,\, \Omega=\sqrt{G\phi} \,, \label{confo1}\\
&&\nonumber \\
&& d\tilde{\phi}=\sqrt{ \frac{\left| 2\omega(\phi)+3 
\right|}{16\pi G}} \, 
\frac{d\phi}{\phi} \label{confo2}
\end{eqnarray}
for $\omega\neq -3/2$, the scalar-tensor action~(\ref{STaction}) 
assumes the Einstein frame form
\begin{eqnarray}
&& S_{ST}  =  \int d^4x \, \sqrt{-\tilde{g}} \left[
\frac{\tilde{R}}{16\pi G} -\frac{1}{2}\, \tilde{g}^{ab} 
\tilde{\nabla}_a \tilde{\phi} 
\tilde{\nabla}_b \tilde{\phi} 
-U(\tilde{\phi}) \right. \nonumber\\
&& \left. +     \frac{ {\cal L}^{(m)} }{ 
(G\phi)^2} \right]  \,,
\end{eqnarray}
where a tilde denotes quantities associated with the 
rescaled metric $\tilde{g}_{ab}$, ${\cal L}^{(m)}$ is the matter 
Lagrangian density, and 
\be \label{questaA}
U\left( \tilde{\phi} \right)=\frac{V\left[ \phi( \tilde{\phi}) 
\right]}{\left[ G\phi ( \tilde{\phi}) \right]^2} \,.
\ee 
This is formally the action of  General Relativity with a 
minimally coupled scalar field $\tilde{\phi}$ but with the 
important difference 
that this scalar now couples explicitly to matter and the units 
of time and length scale with $\Omega$, while the unit of mass 
scales as 
$\Omega^{-1}$ \cite{Dicke}. The 
Einstein frame field equations are
\begin{eqnarray}
&&  \tilde{R}_{ab}-\frac{1}{2}\, 
\tilde{g}_{ab}\tilde{R}=\frac{8\pi G}{(G\phi)^2}\, T_{ab}^{(m)} +
8\pi G \, \tilde{T}_{ab}^{(\tilde{\phi})} \,,\\
&&\nonumber\\
&& \tilde{\Box} \tilde{\phi} -\frac{ 
dU}{ d\tilde{\phi} }=\frac{8\pi G T^{(m)} }{(G\phi)^2} 
\,,\label{Eframeeqforscalar}
\end{eqnarray}
whereas 
\be
\tilde{ T}_{ab}^{(\tilde{\phi})}=\tilde{\nabla}_a \tilde{\phi}
\tilde{\nabla}_b \tilde{\phi} - \frac{1}{2}\, \tilde{g}_{ab} 
\tilde{g}^{cd}\tilde{\nabla}_c \tilde{\phi}
\tilde{\nabla}_d \tilde{\phi} -\frac{U\left( \tilde{\phi} 
\right)}{2} \, \tilde{g}_{ab}  
\ee
is the canonical stress-energy  tensor for a scalar 
field minimally coupled  with the curvature, which satisfies the 
weak energy condition if $V\geq 0$. If the metric $g_{ab}$ is of 
the spherically symmetric  
form~(\ref{metric}) also the rescaled 
$\tilde{g}_{ab}$ assumes the same form with 
$\Omega=\Omega(\phi)=\Omega(t,r)$. 

It is well known that, at the classical level, the Jordan and 
the Einstein conformal frames are equivalent descriptions of the 
same theory   \cite{Dicke, Flanagan, FaraoniNadeauconfo} when the 
conformal transformation is 
well-defined, and one 
can recover  easily in the Einstein frame 
the results discussed in the previous sections.
When the scalar field $\tilde{\phi}$ is constant (which only 
happens if its Jordan frame cousin $\phi$ is constant) then one 
obtains the same equations of motion as in General Relativity 
with a cosmological constant (if $U( \tilde{\phi}) \neq 0$, which 
is equivalent to $V(\phi) \neq 
0$ and $\phi \neq 0$, as follows from eq.~(\ref{questaA})), and 
version~1 of the 
Jebsen-Birkhoff theorem is recovered.

If $ \phi$ is not constant but is assumed to be 
independent of the time coordinate, $\tilde{\phi}$ given 
by eq.~(\ref{confo2}) is static as well and $\Omega=\Omega (r)$. 
Then, introducing the rescaled four-velocity $\tilde{u}^a= 
\frac{u^a}{\Omega}= \left( \tilde{A}^{-1}, 0,0,0, \right)  $, 
it is
\begin{eqnarray}
\rho [ \tilde{\phi}] & \equiv & 
\tilde{T}_{ab}^{\tilde{(\phi})} \, \tilde{u}^a \, \tilde{u}^b    
=\frac{ \tilde{\phi}'^2  
}{2\tilde{B}^2}  +\frac{ U \left( \tilde{\phi} \right)}{2} \,,
\label{rhophitilde} \\
&&\nonumber\\
J_{(r)} [ \tilde{\phi}] & \equiv & 
- \tilde{T}_{ab}^{\tilde{(\phi})} \, \tilde{u}^a \, 
\tilde{e}^b_{(r)}  = 0 \,, \\
&&\nonumber\\
P_{(r)} [ \tilde{\phi}] & \equiv  & 
\tilde{T}_{ab}^{\tilde{(\phi})} \, \tilde{e}^a_{(r)} \,   
\tilde{e}^b_{(r)}  = \frac{\tilde{\phi}'^2}{2\tilde{B}^2} 
-\frac{U(\tilde{\phi})}{2} \,, \label{Pphitilde} \nonumber\\
&& 
\end{eqnarray}
in the Einstein frame. By assuming that $\phi$ (or, equivalently,  
$\tilde{\phi}$) is 
static, the 
effective energy distribution described by 
$\tilde{T}_{ab}^{(\tilde{\phi})} $ is static and, if 
$T_{ab}^{(m)}$ is 
static as well, version~1 of the Jebsen-Birkhoff theorem holds 
and the spherically symmetric solution of the field 
equations $g_{ab}$ is static in a region in which the coordinate 
gradients preserve their causal character.

In the case $\omega=-3/2$ the field $\tilde{\phi}$ is not defined 
but nothing forbids the use of $\left( \tilde{g}_{ab}, \phi 
\right)$ as Einstein frame variables. The action then becomes
\begin{eqnarray}
&& S_{(-3/2)}  =  \int d^4x \, \sqrt{-\tilde{g}} \left[
\frac{\tilde{R}}{16\pi G} + \frac{3}{2}\, \tilde{g}^{ab} 
\tilde{\nabla}_a  \phi  
\tilde{\nabla}_b \phi  
-V( \phi ) \right. \nonumber\\
&&\nonumber\\
&& + \left.   \frac{ {\cal L}^{(m)} }{ (G\phi)^2} 
\right] \,,
\end{eqnarray}
in which $\phi$ is a phantom field with negative-definite kinetic 
energy, which implies a sign change in the first term on the 
right hand sides of eqs.~(\ref{rhophitilde}) and (\ref{Pphitilde}).  
Again, the metric 
will be static only if $\phi$ and $T_{ab}^{(m)}$ are static and 
the theory reduces to General Relativity with a  cosmological 
constant if $\phi$ is constant.

It must be emphasized that the equivalence between Jordan and 
Einstein frames holds only when the conformal transformation is 
well-defined and breaks down if $\phi \rightarrow 0^{+}$ or 
$\phi \rightarrow +\infty$, which can happen approaching a black 
hole horizon. For example, the black hole solutions of Brans 
class~I \cite{Brans}, those of Bekenstein \cite{Bekenstein}, 
and those of Campanelli and Lousto  
\cite{CampanelliLousto} are spherically symmetric and static but 
they are not Schwarzschild. They cause an apparent contradiction 
with a theorem by Hawking which, loosely speaking, states that 
stationary black  holes in Brans-Dicke theory are the same as the 
stationary black  holes of General Relativity and employs the 
Einstein frame in its proof. This apparent contradiction has 
generated some confusion in the literature and is discussed in 
the next subsection.

\subsection{Hawking's theorem and the  Jebsen-Birkhoff theorem in 
Brans-Dicke gravity}

Hawking's theorem \cite{HawkingBD} states that  a 
stationary spacetime containing a black hole is a solution of the  
Brans-Dicke field equations (with  $V=0$) if and only if it is a 
solution of the Einstein field equations, and therefore 
it must be axially symmetric or static.  The theorem is usually 
taken to mean that Brans-Dicke black holes  are exactly the same 
as those of General Relativity: 
this is an overstatement and in fact many solutions of 
scalar-tensor theories including, but not necessarily 
limited to, Brans-Dicke gravity are known which 
describe black holes with a static scalar field and do not 
coincide with the Schwarzschild metric. For these solutions the 
scalar field either goes to zero or diverges on an event 
or apparent horizon and 
this feature invalidates the proof of Hawking's theorem, as we 
shall see 
 below. The most well known example is probably that of Brans' 
class~I solutions given by \cite{Brans}
\begin{widetext}
\begin{eqnarray}
ds^2 & = & -  \left(  \frac{ 1-\mu/r}{1+\mu/r}  
\right)^{2/\lambda} dt^2+ \left( 1+\frac{\mu}{r} \right)^4 \left( 
\frac{1-\mu/r}{1+\mu/r} \right)^{ \frac{ 2\left( \lambda -C -2 
\right)}{\lambda} }\left( dr^2+r^2 d\Omega_2^2 \right) \,,
\label{BransImetric} \nonumber\\
&&\\
\phi(r) &=& \phi_0 \left( \frac{1-\mu/r}{1+ \mu/r} 
\right)^{C/\lambda} 
\,,  \label{BransIscalar}
\end{eqnarray}
\end{widetext}
where $ \mu ,C,\phi_0$ and $\lambda$ are constants with  
\be
\lambda^2=\left( C+1 \right)^2 -C\left( 1-\frac{\omega C}{2} 
\right) >0 \,.
\ee
Clearly, this metric is static and not the Schwarzschild solution 
while  it satisfies  the vacuum Brans-Dicke field equations with 
$V=0$ for  $r> \mu $ (three other classes of spherically 
symmetric solutions were found 
by Brans \cite{Brans}, although they are not  
all independent from each other \cite{BhadraSarkar}). 
Note that for the positive values of $C$ and $\lambda$ usually 
considered, the Brans-Dicke scalar field goes to zero on the 
horizon.

Let us consider now the proof of Hawking's theorem, which is 
performed in the Einstein frame \cite{HawkingBD}.  Another  
theorem \cite{Hawkingprevious} states that a 
stationary black hole in General Relativity must be axisymmetric 
and have spherical topology and relies on the weak (or the null) 
energy condition being satisfied. Hawking's theorem extends 
this previous theorem to Brans-Dicke 
theory \cite{HawkingBD}, aiming to prove that the scalar 
field 
is static. It is pointed out in \cite{HawkingBD} that  the 
advantage of going to 
the Einstein frame is that the rescaled Brans-Dicke scalar 
 \be
\tilde{\phi}= \sqrt{   \frac{\left| 2\omega +3\right| }{16\pi G} 
}  \, \ln \left( \frac{\phi}{\phi_*} \right) 
\ee
(obtained by integrating eq.~(\ref{confo2}), where $\phi_*$ is a 
constant) has canonical kinetic energy  
density and obeys the weak and  null energy 
conditions. The assumption that spacetime is stationary then 
implies that it is also  axially symmetric 
\cite{Hawkingprevious} and, therefore, there exist 
a timelike Killing field $t^a$ and a spacelike Killing 
field $\psi^a$ (outside the horizon) and the Einstein 
frame  scalar  $\tilde{\phi}$ must necessarily be  constant 
along the orbits of $t^a$ and $\psi^a$ 
in order 
to respect these symmetries, hence $\partial^a \tilde{\phi} $ can 
only be spacelike or zero outside the horizon.

Consider now a 4-dimensional 
volume ${\cal V}$ bounded by two Cauchy hypersurfaces ${\cal S}$ 
and $  {\cal S}'$ at two consecutive instants of time,  a portion 
of 
the black hole event horizon, and spatial infinity 
\cite{HawkingBD}: the 
Einstein frame equation of motion~(\ref{Eframeeqforscalar})  in 
vacuo and with 
$V=0$ becomes $\tilde{\Box} \tilde{\phi}=0$, where 
$ \tilde{\Box} \equiv \tilde{g}^{ab}\tilde{\nabla}_a 
\tilde{\nabla}_b $ is 
the Einstein frame  d'Alembertian. Multiplying this equation by 
$\tilde{\phi}$, integrating over ${\cal V}$, and using the Gauss 
theorem and the identity  $ \tilde{\phi} \tilde{\Box} 
\tilde{\phi}= \tilde{\nabla}^c \left( \tilde{\phi} 
\tilde{\nabla}_c \tilde{\phi} \right) - 
\tilde{\nabla}^c \tilde{\phi} 
\tilde{\nabla}_c \tilde{\phi} $,  one obtains 
\cite{HawkingBD}
\be
\int_{ {\cal V} } d^4x \, \tilde{g}^{ab} 
\tilde{\nabla}_a \tilde{\phi} 
\tilde{\nabla}_b \tilde{\phi} = \int_{ \partial {\cal V} } 
dS^c \left( \tilde{\phi}  \tilde{\nabla}_c \tilde{\phi} \right) 
\, .
\ee
The integral over the boundary $\partial {\cal V}$ on the right 
hand side is split into four  contributions: 
\begin{widetext}
\be
\int_{ \partial {\cal V}} 
dS^c \left( \tilde{\phi}  \tilde{\nabla}_c \tilde{\phi} \right) = 
\left( \int_{ {\cal S} } + \int_{ {\cal S}' } + \int_{r=+\infty } 
+ \int_{horizon } \right)  dS^c
\left( \tilde{\phi}  \tilde{\nabla}_c \tilde{\phi} \right) \,.
\ee
\end{widetext}
The contributions from the portions of the Cauchy 
hypersurfaces ${\cal S}$ and $ {\cal S}' $ cancel out because 
they 
have the same absolute value due to the time symmetry but 
opposite signs because of the opposite directions of the 
outgoing unit normal on these hypersurfaces. The contribution 
from spatial infinity vanishes because $\tilde{\phi}$ vanishes 
there, whereas the contribution from the integral over the 
portion of 
the horizon is supposed to vanish because the 
projection of $\partial^a \tilde{\phi} 
$ along the null vector tangent to the horizon, which 
is a linear combination of $t^a$ and $\psi^a$, vanishes due to 
the symmetries \cite{HawkingBD}. This point is crucial: 
the argument is not valid in general because  the Einstein frame 
scalar $\tilde{\phi}$ may not be 
defined at the horizon and, indeed, this is the case for the 
Brans class~I solutions (\ref{BransImetric}) and  
(\ref{BransIscalar}) in which  the Einstein frame scalar 
$\tilde{\phi} 
\propto \ln \phi $ diverges on the horizon because the Jordan 
frame $\phi \rightarrow 0^{+} $  there. The conformal 
transformation to the Einstein frame becomes ill-defined at the 
horizon and its variables $\left( \tilde{g}_{ab}, 
\tilde{\phi} \right)$ cannot be used on this surface.  Of 
course, nothing 
forbids to use the Jordan frame instead of the Einstein one, but 
then  the scalar 
$\phi$ violates the weak and null  energy conditions because its 
stress-energy tensor given by eq.~(\ref{STfield1}) has  a 
non-canonical 
structure  
containing second derivatives of $\phi$ instead of being 
quadratic in the first derivatives (indeed, it is now well known 
that nonminimally coupled scalar fields can violate all the 
energy conditions \cite{ECviolation}).

The scalar field in the known solutions violating the Hawking 
theorem does turn out to be  static (which is what 
\cite{HawkingBD} intended to prove), but 
it has  a radial dependence and these solutions do 
not coincide with the Schwarzschild metric.

Campanelli and Lousto \cite{CampanelliLousto} report   
static solutions of Brans-Dicke theory which possess a static but 
radially-dependent scalar. These authors comment on the violation 
of the null energy condition in the Jordan frame being the cause 
of the violation of the no-hair theorem. However, this comment 
(echoed in \cite{BhadraSarkar}) appears to be a bit misleading 
because Hawking's theorem discusses  
the Einstein frame scalar $\tilde{\phi}$ which does satisfy the 
energy condition, while Campanelli and Lousto refer to the 
Jordan frame $\phi$ which doesn't.  Their comment is 
technically correct but, {\em per se}, does not help 
explaining the violation of Hawking's theorem. Note that, if the 
Brans-Dicke scalar $\phi$ does not go to zero or diverges on the 
horizon, Hawking's theorem applies and the solution is forced to 
be Schwarzschild.

\section{The Jebsen-Birkhoff theorem in $f(R)$ gravity}

We are now ready to come back to $f(R)$ gravity, which is the 
original motivation for our work, even though  the 
understanding of 
the Jebsen-Birkhoff theorem and spherical symmetry in 
scalar-tensor gravity has merit in itself.  The formal 
equivalence of  $f(R)$ theories with scalar-tensor gravity has 
beeen discovered  and rediscovered many times 
\cite{STequivalence}. Metric $f(R)$ 
gravity is equivalent to a Brans-Dicke theory with Brans-Dicke 
parameter $\omega =0$ and a non-trivial potential, while Palatini 
$f(R)$ gravity is equivalent to an $\omega=-3/2$ Brans-Dicke 
theory with potential (see \cite{review} for details). In the 
Palatini version,
the scalar  $\phi$ is non-dynamical, as has been 
pointed out in  various works \cite{BarausseSotiriouMiller, 
review, PalatiniPLB,  Cauchy}. The metric-affine version of 
$f(R)$ gravity \cite{metricaffine} is not equivalent to a 
scalar-tensor theory and will not be considered here.

\subsection{Palatini $f(R)$ gravity}

Let us consider now the $ \omega=-3/2, V\neq 0$ 
Brans-Dicke equivalent of Palatini $f(R)$ gravity. In vacuo, 
electro-vacuo, or in any region in which the trace of the matter 
energy-momentum $T^{(m)}$ is constant, the d'Alembertian 
disappears from the field 
equation~(\ref{STfield2}), which reduces to 
\be\label{Palatinialgebraic}
8\pi G T^{(m)}+ \phi \, \frac{dV}{d\phi}-2V(\phi) =0 \,,
\ee
no longer a  differential but an algebraic or 
trascendental equation.   If eq.~(\ref{Palatinialgebraic})  has 
solutions they are of the form $\phi=$const.$\equiv \phi_0$. 
In this case the  field equation~(\ref{STfield1}) reduces to 
\be
R_{ab}-\frac{1}{2}\, g_{ab}R  = \frac{8\pi}{\phi_0} \, 
T_{ab}^{(m)} - \frac{V(\phi_0)}{2\phi_0} g_{ab} 
\,,
\ee
which describes General Relativity with a cosmological constant 
$ \Lambda=\frac{V(\phi_0)}{2\phi_0}$, 
for which Birkhoff's theorem holds if the matter distribution 
described by $T_{ab}^{(m)}$ is static (including the vacuum case 
$ T_{ab}^{(m)}=0$) which is consistent, of course, with the 
previous assumption $T^{(m)}=$const.

To summarize, the Jebsen-Birkhoff theorem holds in Palatini 
$f(R)$ gravity with a static matter distribution; the fact that 
the Jebsen-Birkhoff theorem holds when $T_{ab}^{(m)}=0$ is well 
known \cite{review}  and is due to the 
non-dynamical nature of the Brans-Dicke scalar present in this 
class of theories, which acts as an effective form of matter 
without dynamics (this feature has been studied in  
detail, see the discussion and  references in \cite{review}).

\subsection{Metric $f(R)$ gravity}

Metric $f(R)$ gravity is equivalent to an $\omega=0$ Brans-Dicke 
theory with a non-trivial potential. This time, in vacuo, the 
field equation~(\ref{STfield2}) reduces to
\be
\Box \phi = \frac{1}{3} \left[ \phi \, \frac{dV}{d\phi} -2V(\phi) 
\right] \,,
\ee
which is now a true dynamical equation. Since $\phi$ is 
dynamical and, in general, time-dependent the 
Jebsen-Birkhoff theorem is not 
valid, which has been noted on several occasions 
(see \cite{review} 
and the references therein). Most studies in the literature 
impose the condition $\phi=$const. (equivalent to $R=$const. in 
the original $f(R)$ theory) for ease of calculation, and compare 
static solutions with Solar System experiments.   This point 
of view carries the risk of not exploring the richer variety of 
solutions with $\partial \phi/\partial t \neq 0$, which are 
certainly more generic than static ones.

\section{Conclusions}

Motivated by the recent attention to metric and Palatini $f(R)$ 
gravity theories revived to explain the cosmic acceleration 
without dark energy, we have considered spherical symmetry and 
the Jebsen-Birkhoff theorem in these theories and, by extension, 
in general scalar-tensor gravity.

Generalizing the Jebsen-Birkhoff theorem of General Relativity 
to situations with  matter present allows one to understand 
the validity, or lack thereof, of this theorem in scalar-tensor 
gravity because the  scalar-tensor field equations can be 
rewritten as effective  Einstein equations with the Brans-Dicke 
like scalar acting as  a form of effective matter.  Using the 
Jordan frame description of scalar-tensor gravity,  this  
effective matter distribution must be static in order for the 
Jebsen-Birkhoff theorem to be valid. This conclusion is not 
hard to obtain  but is seems that it is necessary to 
formulate it explicitly in order to make progress with $f(R)$ 
gravity. 
The situation can be summarized as follows: if $\phi$ is static 
the spherically symmetric solution is locally static between 
horizons  
but not necessarily  Schwarzschild-(anti-)de Sitter; if $\phi$ is 
constant the  solutions is Schwarzschild-(anti-)de Sitter.

Since scalar-tensor gravity admits an Einstein frame description 
in which the scalar has canonical form except for the fact that 
it couples directly to matter, the result obtained in the Jordan 
frame must be recovered in the Einstein frame description, and we 
checked that this is indeed the case. The equivalence (at 
the classical level) between Jordan and Einstein frame breaks 
down when the conformal transformation (\ref{confo1}) and 
(\ref{confo2})  becomes ill-defined, and this occurrence allows 
one to 
understand the apparent contradiction between certain 
spherical solutions and Hawking's theorem on Brans-Dicke black 
holes. Shedding light onto this riddle certainly does not have 
deep new consequences (these non-Schwarzschild  static solutions  
have now been known for a long time) but we are not aware of  
an explicit explanation in these terms in the literature.

Once the role of the Jebsen-Birkhoff theorem in scalar-tensor 
gravity is established, it is straightforward to understand that 
the validity of this theorem in Palatini $f(R)$ gravity is yet 
another manifestation of the non-dynamical character of the 
Brans-Dicke scalar present in this theory. Similarly, the failure 
of the theorem in metric $f(R)$ gravity reflects the dynamical 
nature of the scalar degree of freedom present in these 
theories. Most current studies of spherically symmetric solutions 
in metric $f(R)$ gravity focus on static solutions missing 
time-dependent solutions which are, without doubt, more generic 
than static ones (although there is at present no mathematically 
well-defined meaning of ``generic''). To complicate the 
issue, metric $f(R)$ theories of current interest are designed 
to produce an effective time-varying cosmological constant in 
order to explain the present acceleration of the universe without 
dark energy, and it is expected that ``generic'' solutions (if a 
meaning can be assigned to this adjective) will be asymptotically 
Friedmann-Lemaitre-Robertson-Walker solutions violating  the 
Jebsen-Birkhoff theorem. Some solutions of this kind are known in 
General Relativity \cite{McVittie, cosmobhs}, scalar-tensor 
gravity \cite{CliftonMotaBarrow} and in metric 
$f(R)$ gravity \cite{CliftonCQG}, but they are still not 
understood very well even in the context of General Relativity 
and it will be interesting to study them further in the future.

\begin{acknowledgments}
It is a pleasure to thank Vincenzo Vitagliano for  a discussion 
and  the Natural Sciences and  Engineering Research 
Council of Canada  (NSERC) for financial support.
\end{acknowledgments}


\end{document}